# Does human activity widen the tropics?

Katya Georgieva and Boian Kirov

*Solar-Terrestrial Influences Laboratory at the Bulgarian Academy of Sciences*

The progress article "Widening of the tropical belt in a changing climate" by Seidel et al. (2008) published in the first issue of Nature Geosciences, summarizes the results of several methods to determine the width of the tropical zone. All they show evidence that the tropics have been expanding over the past few decades. We confirm this widening based on one more indicator - the position of the subtropical centers of high pressure. However, we question the implication of the authors that the tropics widen in response to human activity, and suggest as a more probable cause the increasing solar activity. Consequently, we question their conclusion that "this widening may continue into the future in association with anthropogenic climate change", and suggest that whether the tropics will continue widening will depend on the future evolution of solar activity rather than on anthropogenic activity.

The subtropical high atmospheric pressure centers form at the descending branches of the Hadley circulation cell, therefore their position is an indicator of the extent of the tropical zone. Both Azores High in the Atlantic and Hawaiian High in the Pacific whose position and intensity are calculated since 1899 (Hameed et al., 1995), have moved poleward since then, confirming that the tropics have expanded over the past century. In this period, solar activity has increased, and its long-term variations are highly correlated to the long-term variations in the latitudes of the two centers of action (Kirov and Georgieva, 2005). Solar activity can be decomposed (Feynman, 1982) into two parts: one proportional to the solar closed magnetic flux, and hence to the sunspot number and to the intensity of solar electromagnetic radiation, the other proportional to the solar open flux, and to the velocity and magnetic field intensity of the solar wind reaching the Earth. The two are not independent because they are both manifestations of the solar magnetic activity, but their long-term variations are not identical (Georgieva and Kirov, 2006). There are at least two possible mechanisms, probably operating concurrently, linking the variations in these two manifestations of solar activity to the variations in the extent of the Hadley cell. The first accounts for variations in solar electromagnetic radiation. With increasing solar activity, the total irradiance emitted by the Sun increases little, but substantially increases its shorter wavelength fraction which is entirely absorbed in the stratosphere. Model predictions confirmed by analyses of observations described e.g. by Haigh et al.(2005) (cited in the paper by Seidel et al., 2008), show that low-latitude stratosphere heating due to the absorption of more short-wave solar radiation when the Sun is more active, leads to weakening and expansion of the Hadley cells. The second mechanism (Tinsley, 2000) links the changes in the global electric circuit due to variations in the galactic cosmic rays flux, solar energetic particles, relativistic electrons, and global ionospheric potential, all caused by variations in solar open magnetic flux, to changes in cloud cover, atmospheric temperature, pressure distribution and therefore dynamics of the atmosphere. The importance of the first mechanism should depend on geographic latitude (stronger influence for lower geographic latitude), and of the second one – on geomagnetic latitude (stronger influence for higher geomagnetic latitude). Azores high and Hawaiian high are at the same geographic latitude, but their geomagnetic latitudes differ by almost 10$^o$. Consequently, the long-term variations in the latitude of Hawaiian high correlate more strongly (r=0.98) with the part of solar activity proportional to short-wave radiation than with the part proportional to solar wind intensity (r=0.90), while the correlations are reverse (0.84 for electromagnetic radiation versus 0.95 for solar wind intensity) for Azores high. The compliance with the expected differences of the correlations and their high statistical significance (better than 99% in all cases) make us feel that attributing the widening of the tropical zone to human activity before fully accounting for possible natural factors, is premature at the very least.

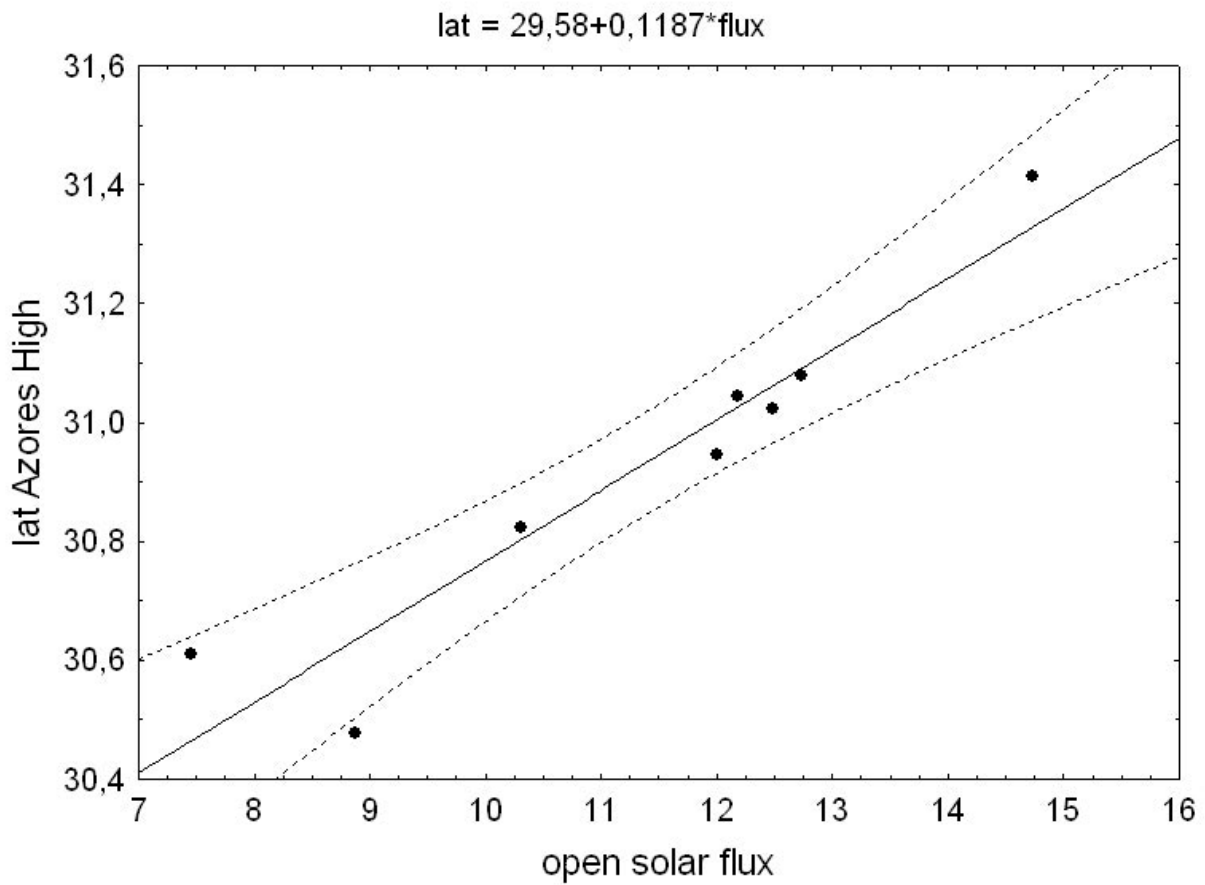

An example: Dependence of the winter latitude of the Azores High on the part of the solar activity proportional to the open magnetic flux (estimated as in Feynman, 1982); climatic normals (30-point moving averages with a step of 10 years, i.e. 1901-1930, 1921-1940, etc.), see Guttman (1989). The dashed lines denote the 95% confidence level.